\documentstyle[epsf,twocolumn,aps]{revtex}
\tightenlines

\newcommand{\be}{\begin{equation}}
\newcommand{\ee}{\end{equation}}
\newcommand{\ba}{\begin{array}}
\newcommand{\ea}{\end{array}}

\newcommand{\bac}{\begin{array}{c}}
\newcommand{\bal}{\begin{array}{l}}
\newcommand{\baR}{\begin{array}{r}}
\newcommand{\bacc}{\begin{array}{cc}}
\newcommand{\ball}{\begin{array}{ll}}
\newcommand{\balr}{\begin{array}{lr}}
\newcommand{\barl}{\begin{array}{rl}}
\newcommand{\baccc}{\begin{array}{ccc}}
\newcommand{\barcl}{\begin{array}{rcl}}
\newcommand{\balcl}{\begin{array}{lcl}}
\newcommand{\barcll}{\begin{array}{rcll}}
\newcommand{\barll}{\begin{array}{rll}}
\newcommand{\barrclcl}{\begin{array}{rrclcl}}
\newcommand{\bacl}{\begin{array}{cl}}
\newcommand{\bacll}{\begin{array}{cll}}
\newcommand{\eac}{\end{array}}
\newcommand{\ber}{\begin{eqnarray}}
\newcommand{\eer}{\end{eqnarray}}

\newcommand{\mnval}{Mn$_2$VAl}

\begin{document}
\twocolumn[\hsize\textwidth\columnwidth\hsize\csname @twocolumnfalse\endcsname



\title{Half-Metallic Ferrimagnetism in \mnval} 

\author{Ruben Weht and Warren E. Pickett}

\address{Department of Physics, University of California, Davis CA 95616}

\date{\today}
\maketitle

\tightenlines

\begin{abstract}
We show that \mnval ~is a compound for which the generalized 
gradient approximation (GGA) to the exchange-correlation functional
in density functional theory makes a qualitative change in 
predicted behavior compared to
the usual local density approximation (LDA).  Application of GGA leads
to prediction of \mnval ~being a half-metallic ferrimagnet, with the
minority channel being the conducting one.  The electronic and magnetic
structure is analyzed and contrasted with the isostructural
enhanced semimetal Fe$_2$VAl.
\end{abstract}
\vskip 1cm
\footnoterule



\newpage

]

\section{Introduction}
The cubic Heusler structure class X$_2$YZ of intermetallic compounds
provides a great variety of behavior, including an interesting variety of
magnetic phenomena.  A striking example is the prediction, solely from
band theory, of half-metallic (HM) ferromagnetism (FM) in this
class (and the related half-Heuslers XYZ).\cite{groot}

In the Heusler systems HM FM has never been unambiguously confirmed by
experiment, although there is strong evidence in some cases.  Only
in the ``colossal magnetoresistance'' system La$_{1-x}$Sr$_x$MnO$_3$ 
have HM signatures been seen directly and clearly in spin-polarized
photoemission experiments,\cite{park} as was predicted from band
theory calculations utilizing the local density approximation 
(LDA)\cite{wepdjs} for the exchange-correlation energy functional. 
Probably several Heusler and half-Heusler compounds are HM.

Over the past fifteen years many (probably at least twenty) HM FM
compounds have been predicted based on LDA, and there has not been
much reason to doubt the computational results 
(for example, these are
not strongly correlated materials).  Recent 
examples include Sr$_2$FeMoO$_6$,
which appears to be supported by experimental data,\cite{tera} and
possible HM materials with vanishing moment -- 
``HM {\it anti}ferromagnets'' -- which have not yet been made.\cite{hmafm}
However, there is a need to make predictions as robust as possible, and
in the double perovskite structure of Sr$_2$FeMoO$_6$ and the proposed
HM antiferromagnets structural distortions are a likely occurrence and
cloud the forecasts based on an ideal structure.  The Heusler structure 
we consider here are much less inclined to distortion,\cite{distort}
leaving the prediction based solely on the quality of the electronic
structure description.

Here we present a case of a Heusler structure compound
where the use of the generalized
gradient approximation (GGA)\cite{gga} to the exchange-correlation
energy functional leads to the prediction of a qualitatively different 
type of behavior compared to
LDA.  GGA has a stronger formal foundation,  because it
accounts specifically for density gradients that are neglected
in LDA, and does so in a way that satisfies several exact constraints
on the form of the exchange-correlation energy functional.  
Put into practice, GGA seems to give a general improvement in comparison
to experimental data for alkali metals,\cite{alkalis} $3d$ and $4d$
transition metals,\cite{3d4d} lanthanides,\cite{lanthan} and ionic
insulators.\cite{ionic}  For covalent semiconductors the reports
are somewhat inconclusive,\cite{covalent} but overall GGA seems to be
roughly as good as LDA.  Assessing the differences between GGA and LDA
is delicate, requiring in principle a full potential method.
Because the GGA potential requires a more demanding
calculation than does LDA, there are not yet many tests for multicomponent
intermetallic compounds.

The compound in question here
is the Heusler compound \mnval.  This compound is somewhat peculiar
regardless of its detailed electronic structure: 
whereas there are many Heusler compounds of the form
X$_2$YZ where Y=Mn, this compound is the only one reported where the
X site is occupied by Mn.  In the Y site Mn has high spin (around
4 $\mu_B$ in X$_2$MnZ compounds\cite{mub4}), while in the
X site in \mnval ~it has been found in LDA calculations 
that it has a low spin, in agreement with measurements.

Although
V and Al atoms may substitute on each other's sublattice, in the system
Mn$_2$V$_{1+x}$Al$_{1-x}$ the lattice constant and xray intensities
show a kink at $x$=0, clearly identifying the stoichiometric 
composition.\cite{yoshida} The structure remains the Heusler one
from $x$ between -0.3 and +0.2, with linearly varying saturation moment.
At stoichiometry the saturation moment is reported to be
1.9 $\mu_B$, close to the
integral value characteristic of the spin moment of
HM magnets,\cite{groot,rudd} and the Curie
temperature is T$_C$ = 760 K.  The Curie constant obtained from high
temperature susceptibility (above T$_C$) is consistent with the 
saturation moment.  Nakamichi and Stager\cite{naka} inferred from
NMR spin echo data that Mn and V have moments near 1.2 and -0.7 $\mu_B$
respectively, making the net value (2$\times$1.2-0.7)
roughly consistent with the
reported saturation magnetization.  Itoh {\it et al.}\cite{itoh} obtained
from neutron diffraction 1.5$\pm$0.3 and -0.9 $\mu_B$ for Fe and V
respectively.

Another reason for interest in this compound is the recent excitement
caused by the related isostructural compound Fe$_2$VAl.  From heat capacity,
resistivity, and photoemission data this compound appears to be 
`chubby fermion' metal, while LDA (and GGA) 
calculations\cite{feval1,feval2,feval3}
indicate that it has a semimetallic band structure with
a very small number of carriers.  One likely scenario is that its 
effective mass is enhanced by dynamic electron-hole correlations.
\cite{feval3} 

In this paper we provide the results of accurate GGA calculations for
\mnval.  LDA calculations were presented earlier by Ishida, Asano,
and Ishida\cite{ishida}, who obtained a near-HM situation that is
reproduced by our LDA calculations.  The important feature is that 
GGA predicts a HM FM (actually, ferrimagnetic, due to the moment of
V that opposes the Mn moments).  Recently there is increasing evidence
that very large magnetoresistance and half-metallicity are closely
related, for example in the `colossal magnetoresistance' 
manganites\cite{park}
and in the double perovskite compound Sr$_2$MoRuO$_6$.\cite{tera}
Due to this apparent close connection
between HM behavior and large magnetoresistance, our predictions
suggest that the magnetoresistance of \mnval ~should be measured.
In Sec. II we describe the structure and outline our method of
calculation.  The results and comparison with Fe$_2$VAl are given in
Sec. III.  The conclusions are summarized in Sec. IV.

\section{Heusler Structure, and Method of Calculation}
Although the Heusler structural class includes an alloy 
system where the atoms on the Y and Z
sites often intermix, indications are that at stoichiometry this 
compound corresponds to
an ideal Heusler (L2$_1$)
structure compound.  This structure type, pictured in Fig. 1, is
based on an underlying bcc arrangement of atomic sites with lattice constant
$a/2$, with V at (0,0,0),
Al at ($\frac{1}{2},\frac{1}{2},\frac{1}{2})a$, and Mn atoms at
($\frac{1}{4},\frac{1}{4},\frac{1}{4})a$ and
($\frac{3}{4},\frac{3}{4},\frac{3}{4})a$, where $a$=5.875~\AA~is
the lattice constant of the resulting fcc compound.

This structure is also that of Fe$_2$VAl and of Fe$_3$Al, 
the latter of which has two inequivalent Fe sites
(X and Y sites).  There is no report that Mn$_3$Al forms in this structure.

\begin{figure}[tbp]
\epsfxsize=4.5cm\centerline{\epsffile{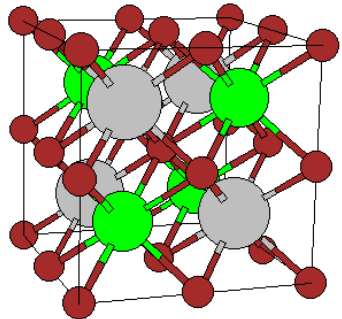}}
\caption{The crystal structure of the Heusler compound Mn$_2$VAl.
The small spheres denote the Mn sites, the medium and large spheres
are the V and Al sites.  The volume shown includes four primitive cells.
\label{Fig1}}
\end{figure}

We have applied the linearized augmented plane wave method\cite{djsbook}
that utilizes a fully general shape of density and potential. 
The WIEN97 code\cite{wien} has been used in
the calculations.  The lattice constant of 5.875~\AA~was used.
LAPW sphere radii (R) of 2.00 a.u. were chosen
with cutoffs of RK$_{max}$ up to 8.7, providing well
converged basis sets with more than
500 functions per primitive cell.
Self-consistency was carried out on k-points meshes of around 200
points in the irreducible Brillouin zone
(12$ \times12\times 12$ and $20 \times 20 \times 20$
meshes).  The GGA exchange-correlation functional of Perdew {\it et al.}
\cite{gga} was used in the present work.

\section{Discussion of Results}
\subsection{Ferrimagnetic \mnval}

The majority and minority (which we will refer to as up and down) 
band structure resulting from GGA
are shown in Fig. 2.  These band structures are similar to what is 
obtained using LDA (see Ishida {\it et al.}\cite{ishida}) except for the
important difference that the gap in the majority bands at the Fermi
level E$_F$ is increased by 40\% (from 0.25 eV to 0.35 eV).  The
position of E$_F$ is essentially pinned by the minority band filling,
and there is not any comparable band shifts in the minority bands
within a few tenths of eV of E$_F$.  The result of GGA, arising from
the increase in the gap and the associated band rearrangements, is to 
empty the valence holes in the majority bands.  The net (spin) magnetic moment
is exactly 2 $\mu_B$ per unit cell.  The magnetic alignment is ferrimagnetic,
with roughly a moment of 1.5 $\mu_B$ on each Mn and --0.9 $\mu_b$ on
V (see below for more discussion).

\begin{figure}[tbp]
\epsfxsize=8.0cm\centerline{\epsffile{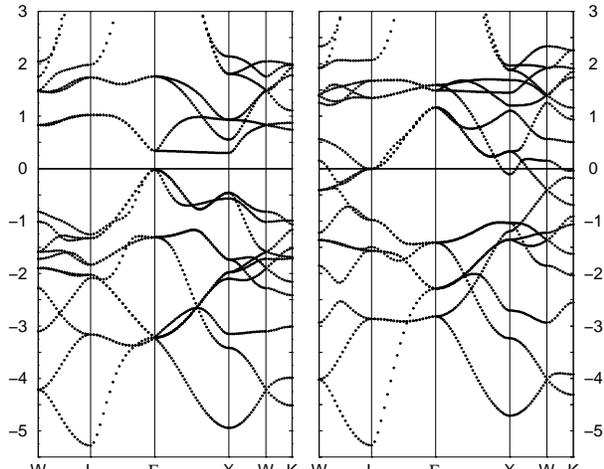}}
\caption{Band structure of half-metallic \mnval: left panel, majority;
right panel, minority.  Note the extreme dissimilarity of the two sets
of bands between -1 eV and 2 eV.  One band, which is largely Al, 
lying below these bands in the -9 eV to -6 eV range, is
not shown.
\label{Fig2}}
\end{figure}

A very noticeable occurrence is just how different the majority and
minority bands, shown in Fig. 2, are within
1 eV of E$_F$.  In the majority bands the direct
bandgap of 0.36 eV occurs at $\Gamma$; in the minority bands the nearest
bands to E$_F$ are 1 eV away at $\Gamma$,
both higher and lower.  This difference,
which is a drastic departure from simple Stoner exchange splitting,
is related to ferrimagnetism, for which the exchange potentials are
of opposite signs on the Mn and V atoms.

\begin{figure}[tbp]
\epsfxsize=7.8cm\centerline{\epsffile{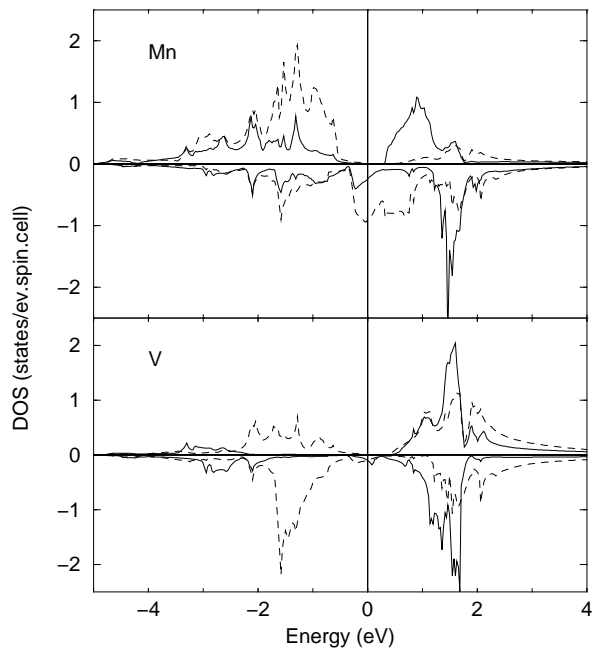}}
\caption{Atom and symmetry projected density of states of \mnval,
with majority plotted upward and minority plotted downward.  
$e_g$ and $t_{2g}$ symmetries are solid and dashed lines, respectively.
\label{Fig3}}
\end{figure}

The atom and symmetry projected densities of states (PDOS) shown in
Fig. 3 clarify the characters of the bands.  The spin up bands below
the gap are $\approx$80\% Mn, including all of the majority Mn $t_{2g}$
states.  Hence the Mn moment is strongly $t_{2g}$ in character.  Since
for V only $t_{2g}$ states are occupied in either spin channel, the
V moment is also $t_{2g}$ in character.  The Al character is so
small as to be difficult to see on the scale of Fig. 3, so it is not shown.

Above the gap lies most of the V spin up states, along with
much of the Mn $e_g$ states.  In the spin down states below 0.5 eV,
there are roughly equal amounts of V ($t_{2g}$) and Mn (mixed) states.
Around E$_F$, however, the DOS is overwhelmingly Mn (strongly $t_{2g}$)
in character.  Hence the (100\% polarized) charge carriers are 
associated almost entirely with Mn atoms.

The mechanism for antiparallel alignment of the V and Mn spins is
not evident from our calculations.  When parallel moments on V and
Mn are used to begin the calculation, the V spin flips direction.
This behavior suggest the FM alignment is substantially above the
ferrimagnetic one in energy.
Although it is tempting to speculate that the opening of the gap
in the majority channel lowers the energy, which would provide a
specific driving force toward half-metallicity, there is no particular
evidence from our calculations to support such a scenario.

\subsection{Fermi Surface}
With the exception of positron annihilation studies of the 
half-Heusler compound NiMnSb,\cite{ACAR}, Fermi surfaces
of half metallic ferromagnets have not been directly measured,
so we provide a brief description of the Fermi surface.
The Fermi surface, which is of course solely for the minority bands,
is rather interesting.  It consists of
small X-centered electron ellipsoids (nearly spherical) containing
electrons, a large ``jungle gym'' type surface with arms along the
cubic axes (Fig. 4), and a highly multiply-connected sheet that might
be considered as centered at the L point of the Brillouin zone
(Fig. 5).  From Fig. 2, it can be seen that there is a near
degeneracy of three bands (within 12 meV)
that fall almost exactly at (actually straddle) the Fermi level.

\begin{figure}[tb]
\epsfxsize=5.2cm\centerline{\epsffile{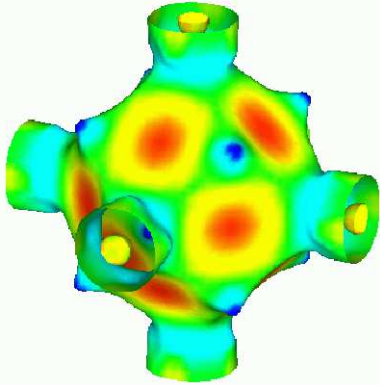}}
\caption{Fermi surface of the lower band, which is of the general
``jungle gym'' type centered at $\Gamma$.  
There is also a tiny spheroid
centered at the X points.  The shading reflects the carrier velocity,
which is smallest along the (111) directions and largest along the
(110) directions.
\label{Fig4}}
\end{figure}

\begin{figure}[tb]
\epsfxsize=5.2cm\centerline{\epsffile{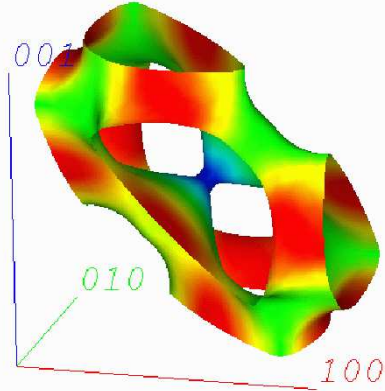}}
\caption{Fermi surface of the higher band, with the cubic axes
provided for orientation.  This furface is multiply connected, with
the L point (center of the hexagonal face) lying at the center of
the portion that is shown (in one octant of the zone).  The narrow neck
at the L point is very sensitive to the Fermi level position.
\label{Fig5}}
\end{figure}

This coincidence leads to identifiable structure in both Fermi
surfaces.
The jungle gym sheet has protrusions along the (111) direction very near
the L point, while the surface in Fig. 5 has a narrow neck at the
L point..
The jungle gym also has flat regions perpendicular to the (110) directions
that provide a nesting wavevector at (0.85,0.85,0)$\pi/a$ and
symmetry related wavevectors.  The surface in Fig. 5 has some flattish
regions but no obvious strong nesting features.

\subsection{Comparison with Fe$_2$VAl}
Both from experiment and from the band structure, Fe$_2$VAl 
is vastly different from \mnval ~in spite of its identical structure 
and closely related constituents.  The total and projected DOS of
Fe$_2$VAl is shown in Fig. 4 for comparison with \mnval.  Fe$_2$VAl
is a semimetal, practically a semiconductor, with a very deep and
1 eV wide pseudogap centered on the Fermi level.  In fact, the
majority bands of \mnval, which are isoelectronic with the majority 
bands of Fe$_2$VAl, have rather similar projected DOS, with the 1 eV
pseudogap in Fe$_2$VAl becoming a 1.5 eV pseudogap with a true 
gap in the center for Mn$_2$VAl.  The pseudogap in Fe$_2$VAl no doubt plays 
an important role in stabilizing the non-magnetic state in that compound.

\begin{figure}[tbp]
\epsfxsize=7.8cm\centerline{\epsffile{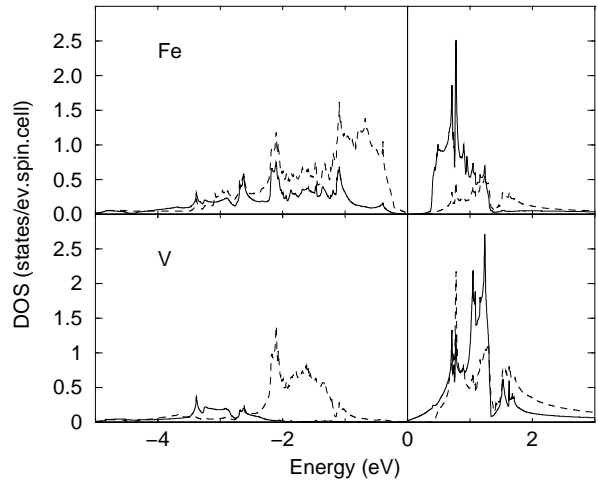}}
\caption{Atom and symmetry projected density of states of Fe$_2$VAl
from Ref. 15,
plotted as in Fig. 3 except that this compound is non-magnetic.  Comparison
with Fig. 3 shows that the majority states of \mnval ~are comparable to
Fe$_2$VAl, while the minority states of Fe/Mn differ strongly.
\label{Fig6}}
\end{figure}

In the minority channel the electronic structures 
of the two compounds are quite distinct.
Where in Fe$_2$VAl there is a pseudogap, in \mnval ~the X site (Mn)
$t_{2g}$ DOS is displaced to higher energies, and forms a partially
occupied band with width $\approx$1.5 eV.  This same Mn $t_{2g}$
character in the majority shifts downward (from the exchange
splitting) and is the cause of opening the majority spin band gap
that leads to HM magnetism.

In Fe$_2$VAl the conduction band states just above E$_F$ are purely
V $e_g$ in character.  In \mnval ~the DOS within nearly 0.5 eV of the
Fermi level, both above and below, are dominated by Mn $t_{2g}$
character.  The main onset of V $e_g$ character has been pushed
nearly 1 eV upward by the exchange splitting and the resulting
changes in bonding.

\section{Summary}
We have shown that the GGA exchange-correlation functional leads to
an increase of 40\% (0.1 eV) in the bandgap in the 
majority bands leading to 
a HM band structure for \mnval, whereas LDA predicted band overlap
in the majority channel.  Since there is no reason to expect 
the structure in the high symmetry Heusler lattice to distort (and
no distortion has been observed), and we
have used the most advanced exchange-correlation functional known at
present, this result represents a robust prediction of modern electronic
structure theory.  Since colossal magnetoresistance has been associated
with the HM state in the ferromagnetic manganites and also apparently 
in the double perovskite Sr$_2$FeMoO$_6$, this compound gives another
clear example to explore the relation between half-metallicity and
large magnetoresistance.
\section{Acknowledgments}
We acknowledge useful communication with E. Kurmaev. 
This work was supported by National Science Foundation Grant DMR-9802076.
Figures 4 and 5 were produced by the AVS graphics package.



\begin{references}
\bibitem{groot}R. A. deGroot {\it et al.}, Phys. Rev. Lett. {\bf 50}, 2024
(1983);
V. Yu. Irkhin and M. I. Katsnel'son, Usp. Fiz. Nauk {\bf 164}, 705 (1994) 
[Sov. Phys. Usp. {\bf 37}, 659 (1994)].
\bibitem{park}J.-H. Park {\it et al.}, Nature {\bf 392}, 794 (1998).
\bibitem{wepdjs}W. E. Pickett and D. J. Singh, Phys. Rev. B {\bf 53},
1146 (1996).
\bibitem{tera}K.-I. Kobayashi, T. Kimura, H. Sawada, K. Terakura, and
Y. Tokura, Nature {\bf 395}, 677 (1998).
\bibitem{hmafm}W. E. Pickett, Phys. Rev. B {\bf 57}, 10613 (1998)
\bibitem{distort}A structural distortion in the antiferromagnetic
Heusler compund Fe$_2$VSi was reported by K. Endo, H. Matsuda, K. Ooiwa
and K. Itoh, J. Phys. Soc. Japan {\bf 64}, 2329 (1995).
\bibitem{gga}J. P. Perdew {\it et al.}, Phys. Rev. B {\bf 46}, 6671 (1992);
J. P. Perdew, K. Burke, and M. Ernzerhof, Phys. Rev. Lett. {\bf 77},
3865 (1996).
\bibitem{alkalis}J. E. Jaffe, L. Zijing, and A. C. Hess, Phys. Rev.
B {\bf 57}, 11834 (1998).
\bibitem{3d4d}K. Kokko and M. P. Das, J. Phys.: Cond. Matter {\bf 10},
1285 (1998); R. Zeller, M. Asato, T. Hoshino, J. Zabloudil, P. Weinberger
and P. H. Dederichs, Phil. Mag. B {\bf 78}, 417 (1998)..
\bibitem{lanthan}A. Delin, L. Fast, O. Eriksson and B. Johansson,
J. Alloys and Compounds {\bf 275-277}, 472 (1998).
\bibitem{ionic}J. K. Dewhurst, J. E. Lowther, and L. T. Madzwara,
Phys. Rev. B {\bf 55}, 11003 (1997).
\bibitem{covalent}C. Filippi, D. J. Singh, and C. J. Umrigar, Phys.
Rev. B {\bf 50}, 14947 (1994); C. Stampfl and C. G. Van de Walle,
Phys. Rev. B {\bf 59}, 5521 (1999); I.-H. Lee and R. M. Martin, Phys.
Rev. B {\bf 56}, 7197 (1997); A. Dal Corso, A. Pasquarello, A. 
Baldereschi, and R. Car, Phys. Rev. B {\bf 53}, 1180 (1996).
\bibitem{mub4}J. K\"ubler, A. R. Williams, and C. B. Sommers, Phys. Rev.
B {\bf 28}, 1745 (1983).
\bibitem{yoshida}Y. Yoshida, M. Kawakami, and T. Nakamichi, J. Phys.
Soc. Japan {\bf 50}, 2203 (1981).
\bibitem{rudd}R. E. Rudd and W. E. Pickett, Phys. Rev. B {\bf 57}, 557 (1998).
\bibitem{naka}T. Nakamichi and C. V. Stager, J. Magn. Magn. Mater.
{\bf 31-34}, 85 (1983).
\bibitem{itoh}H. Itoh, T. Nakamichi, Y. Yamaguchi, and N. Kazama,
Trans. Japan Inst. Metals {\bf 24}, 265 (1983).
\bibitem{feval1}G. Y. Guo, G. A. Botton, and Y. Nishino, J. Phys.: Condens. 
Matter {\bf 10}, L119 (1998).
\bibitem{feval2}D. J. Singh and I. I. Mazin, Phys. Rev. B {\bf 57},
14352 (1998).
\bibitem{feval3}R. Weht and W. E. Pickett, Phys. Rev. B {\bf 58}, 6855 (1998).
\bibitem{ishida}S. Ishida, S. Asano, and J. Ishida, J. Phys. Soc. Japan
{\bf 53}, 2718 (1984).

\bibitem{djsbook}D. J. Singh, {\it Planewaves, Pseudopotentials, and the
LAPW Method} (Kluwer Academic,Boston, 1994).
\bibitem{wien}P. Blaha, K. Schwarz, and J. Luitz, WIEN97, Vienna
University of Technology, 1997. Improved and updated version of the
original copyrighted WIEN code, which was published by P. Blaha,
K. Schwarz, P. Sorantin, and S. B. Trickey, Comput. Phys. Commun.
{\bf 59}, 399 (1990).
\bibitem{ACAR}K. E. H. M. Hanssen and P. E. Mijnarends, Phys. Rev. B {\bf 34},
5009 (1986); K. E. H. M. Hanssen, P. E. Mijnarends, L. P. L. M. Rabou, 
and K. H. J. Buschow, Phys. Rev. B {\bf 42}, 1533 (1990).
\end{references}
\end{document}